\documentclass[twocolumn,showpacs,preprintnumbers]{revtex4}

\usepackage{graphicx}
\usepackage{amsmath}
\usepackage{amssymb}
\def\MF{\mbox{\boldmath$F$}}
\def\MS{\mbox{\boldmath$S$}}
\def\vJ{\mbox{\boldmath$J$}}

\def\MGamma{\mbox{\boldmath$\Gamma$}}
\def\Msigma{\mbox{\boldmath$\sigma$}}

\begin{document}

\title{Metabolic Futile Cycles and Their Functions: 
A Systems Analysis of Energy and Control}
\author{Hong Qian}
\email{qian@amath.washington.edu}
\affiliation{Department of Applied Mathematics, 
University of Washington, Seattle, WA 98195}
\author{Daniel A. Beard}
\affiliation{Biotechnology and Bioengineering Center, 
Department of Physiology, Medical Collge of Wisconsin, 
Milwaukee, WI 53226}

\date{\today}

\begin{abstract}
It has long been hypothesized that futile cycles in cellular
metabolism are involved in the regulation of biochemical pathways.
Following the work of Newsholme and Crabtree, we develop a
quantitative theory for this idea based on open-system
thermodynamics and metabolic control analysis. It is shown that the
{\it stoichiometric sensitivity} of an intermediary metabolite
concentration with respect to changes in steady-state flux is
governed by the effective equilibrium constant of the intermediate
formation, and the equilibrium can be regulated by a futile
cycle.  The direction of the shift in the effective equilibrium
constant depends on the direction of operation of the futile cycle.
High stoichiometric sensitivity corresponds to ultrasensitivity of
an intermediate concentration to net flow through a pathway; low
stoichiometric sensitivity corresponds to super-robustness of
concentration with respect to changes in flux. Both cases
potentially play important roles in metabolic regulation. Futile
cycles actively shift the effective equilibrium by expending energy;
the magnitude of changes in effective equilibria and sensitivities
is a function of the amount of energy used by a futile cycle. This
proposed mechanism for control by futile cycles works remarkably
similarly to kinetic proofreading in biosynthesis. The sensitivity
of the system is also intimately related to the rate of
concentration fluctuations of intermediate metabolites.  The
possibly different roles of the two major mechanisms for cellular
biochemical regulation, namely reversible chemical modifications 
via futile cycles and shifting equilibrium by macromolecular binding, 
are discussed.
\end{abstract}

\maketitle

\section{Introduction}

    Metabolic fluxes and intermediary metabolite
concentrations are the essential currency of intracellular
metabolism.  The fluxes---or rates of turnover of the various
reactions---determine the rate of production of required end
products of a metabolic system, i.e., ATP synthesis by oxidative
phosphorylation and lactate production via glycolysis. The
intermediary metabolite concentrations are often key regulators in
metabolic control.  For example, fructose-1,6-bisphosphate is an
allosteric regulator of pyruvate kinase \cite{BT}. One of the primary 
examples
 of futile cycles in metabolism is fructose-6-phosphate
phosphorylation dephosphorylation, an extensively studied metabolic
regulatory module \cite{selkov,ibarguren,val1,val2}. In the present
work, we study the quantitative relationship between a metabolic
flux and the concentrations of its intermediary metabolites.  In
particular, we study the sensitivity of the intermediate
concentration with respective to changes in the flux.  Depending on
the biological context, the sensitivity may be high for some
metabolites (sensitive) while for others it may be low
(robust)\footnote{This definition of sensitivity is the reciprocal
of the sensitivity of the metabolic flux in a pathway with respect
to changes in the concentration of a metabolite. High sensitivity of
flux to concentration implies robustness in changes in concentration
in response to changes in flux.}. We show that the sensitivity is
intimately related to the energetics, i.e. effective equilibrium
constant, between the intermediate and the upstream substrate(s).
Hence control of sensitivity has a thermodynamic interpretation.
	
    The chemical equilibrium constant is an intrinsic 
property for a given chemical reaction and cannot be
altered by enzyme activity. However, the effective equilibrium
constant operating in a reaction in a network can be shifted via a
futile cycle, as we describe below. This same idea is behind the
concept of kinetic proofreading, in which a kinetic cycle involving
GTP hydrolysis increases the effective binding affinity between a
codon and its tRNA, improving the accuracy of protein biosynthesis
\cite{jjh,ninio,schu}.  At a deeper level, the kinetic proofread
shares a same principle of nonequilibrium physics as the nuclear 
Overhauser effect in magnetic resonance and catalytic wheel \cite{tsong}.

The idea that futile cycles contribute to the regulation of
metabolic functions has a long history
\cite{newsholme,selkov,ibarguren,arkin}. In particular, Newsholme and
Crabtree suggested that futile cycling may be important for
effective regulation at low metabolic fluxes, allowing a pathway to
be controlled with much smaller excursions in the concentrations of
the allosteric regulators. However, there has not been a systematic
mathematical theory substantiating this idea. Because of the lack of
quantitative predictions, experimental verification of this
important idea has been difficult. The present analysis provides
strong support and a mechanistic quantification of the hypothesis of
Newsholme \& Crabtree on the biochemical function of futile cycles
in metabolic systems \cite{newsholme2,newsholme1}. The analysis
focuses on the quantitative relation between the energy expenditure,
thermodynamics, and biochemical control.

\section{Metabolic Flux, Intermediate Concentration,
and Stoichiometric Sensitivity}

\subsection{Sensitivity of Intermediate Concentration to Flux
through a Metabolic Pathway}

    Consider a simple linear metabolic pathway
with species $0$, $1$, and $2$:
\begin{equation}
  0 \overset{k_1}{\underset{k_{-1}}\rightleftharpoons}
  1 \overset{k_2}{\underset{k_{-2}}\rightleftharpoons}
  2 \overset{J}{\longrightarrow}
\label{rxn}
\end{equation}
in which the concentration of the substrate, $c_0$, is fixed. The
constants $k_i$ are either the mass-action rate constants for
first-order chemical kinetics or the effective rate constants acting
around a steady-state. (See Appendix A for a detailed analysis.)
Note that in a metabolic network, there are different ways of
maintaining the input substrate concentration. Two extreme cases are
concentration clamping and constant flux injection \cite{my05bpc}.
While concentration clamping can be achieved experimentally with a
large pool or buffer for the substrate, flux injection is closely
related to the ``flux-generating step'' suggested in
\cite{crabtree}.  Both concentration clamping and flux injection
provide thermochemical driving forces for open biochemical systems,
analogous to batteries in electrical circuits.  

	Recall that there are two types of ideal 
batteries, those that provides constant voltage (voltage 
sources with zero internal resistance) and those that 
provides constant current (current sources with zero internal
conductance). A real battery of course has a finite internal
resistance and conductance.  Metabolic concentration and flux play 
equally important roles in the steady-state of a biochemical system. 
They should be treated on equal footing in a complete theory of 
metabolic dynamics.  Under an ideal setting, one can control the 
concentration(s) and let the fluxes change in response; Similarly,
one can control the flux(s) and let the concentrations change in  
response. Controlling fluxes can, but not necessarily, be accomplished
by changing enzyme activities.  

    Significant controversies exist in the literature on
systems analysis of metabolic networks due to differences in
implicit assumptions on how a system is sustained in a
nonequilibrium steady-state.  In other words, how a system's
steady-state responses to perturbations depends on how the
steady-state is maintained.  See \cite{teusink} for an 
interesting case study.  It is worth pointing out that 
almost all the existing work on metabolic control analysis (MCA) 
implicitly assumes concentration clamping \cite{kacser}.  
In fact, we recently discovered that for systems driven by flux 
injections, the summation of flux control coefficients equals 
zero rather than unity (manuscript in preparation).   
Naturally, the realistic situation in a cell is likely to be mixed. 

    In standard MCA, one defines a flux control coefficient
as the sensitivity of a flux $J$ to an enzyme concentration
\cite{fell1,hs,fell2}. This work, however, focuses on a different
aspect of network control: the steady-state sensitivity of the
concentrations of the intermediates, $c_1$ and $c_2$, to the flux
$J$, at fixed enzyme activities. This stoichiometric sensitivity is
also different from the elasticity coefficient, which is a property
of a single enzyme---an elasticity coefficient (also called local
\cite{kacser} or intrinsic \cite{crabtree}, or immediate \cite{chen}
control coefficient) is determined from the rate law of a single
enzymatic reaction in isolation.  Our stoichiometric sensitivity is
related to the co-response coefficients introduced by Rohwer and 
coworkers \cite{hofmeyr,rohwer}.  The co-response coefficient
emphasizes the concomitantly changes in a flux and a concentration in 
response to a perturbation in a given enzyme.  While changing enzyme 
is one of the possible means to perturb kinetics, there
are other means to perturb a flux.  Hence the stoichiometric
sensitivity is a more general kinetic concept than the co-response
coefficient in MCA.  In fact, the present analysis
focuses on sensitivity of network concentrations to steady-state
network fluxes.  Even though both metabolites and enzymes are
chemical species in a biochemical reaction system, their roles in
metabolic kinetics and thermodynamics are very different. A change
in the enzyme amounts to a change in both forward and backward
effective rate constants for the catalyzed reaction without altering
their ratio (Haldane's equation).

    With given $c_0$ and $J$ as input and output in a steady-state, 
the intermediate concentrations in the reaction of Eq. \ref{rxn} are
\begin{eqnarray}
    c_1 &=& \frac{k_1}{k_{-1}}c_0-\frac{1}{k_{-1}}J
\label{c1}
\\[10pt]
    c_2 &=& \frac{k_1k_2}{k_{-1}k_{-2}}c_0
 - \frac{k_2+k_{-1}}{k_{-1}k_{-2}}J.
\label{c2}
\end{eqnarray}
(See Appendix A.) The stoichiometric sensitivity coefficients are
defined as
\begin{eqnarray}
   \eta_1 =  \left|\frac{\partial\ln c_1}{\partial \ln J}\right|
    &=& \frac{J}{k_{-1}c_1},
\label{eta1}\\[10pt]
    \eta_2 = \left|\frac{\partial\ln c_2}{\partial \ln J}\right|
    &=& \frac{(k_2+k_{-1})J}{k_{-1}k_{-2}c_2} \, .
\label{eta2}
\end{eqnarray}
While the steady-state concentations do not
necessarily increase or decrease in the order from input to output
along the pathway, the sensitivity increases for intermediates as
one moves from ``upstream'' to ``downstream'' ($\eta_2>\eta_1$ since
$k_2c_1-k_{-2}c_2>0$). This observation generally holds true. For a
sequence of reaction of arbitrary length
\begin{equation}
  0 \overset{k_1}{\underset{k_{-1}}\rightleftharpoons}
  1 \overset{k_2}{\underset{k_{-2}}\rightleftharpoons}
  2 \overset{k_3}{\underset{k_{-3}}\rightleftharpoons}
  \cdots
  \overset{k_i}{\underset{k_{-i}}\rightleftharpoons}
  i \overset{k_{(i+1)}}{\underset{k_{-{(i+1)}}}\rightleftharpoons}
  \cdots
  \overset{k_n}{\underset{k_{-n}}\rightleftharpoons}
  n  \overset{J}{\longrightarrow} ,
\label{rxn2}
\end{equation}
the stoichiometric sensitivity for the $i$th intermediate is
\begin{equation}
    \eta_i =  \frac{1 + \frac{k_i}{k_{-(i-1))}}+\cdots
    +\frac{k_i\cdots k_2}{k_{-(i-1)}\cdots k_{-1}}}
    {k_{-i}c_i}J.
\end{equation}
This expression is related to the well-known exit probability and 
mean first passage time out of state $i$ \cite{tk}, as is shown in
\cite{wjh-hq}. The Gibbs free energies of the metabolites, $G_i =
G_i^o+RT\ln c_i$, decreases along the pathway following the
direction of the flux.  As we shall see, this correlation between
the sensitivity and energetics is not just a coincidence.

The sensivity $\eta_1$ can also be expressed as $J/J_-$,
where $J_-$ is the backward flux from state 1 to state 0, and the
net flux $J$ is equal to the foward flux minus the backward flux $J
= J_+ - J_-$.  $\big|\frac{\partial\ln J}{\partial\ln
c_1}\big| = J_- / J$ is an elegant result that was known to Newsholm
and Crabtree \cite{newsholme2}. The flux ratio is a measure of
whether a reaction operates near equilibrium ($J/J_- \ll 1$) or far
from equilibrium ($J / J_- \gg 1$). The expression $\eta_i = J/J_-$
applies to an intermediate at an arbitrary position in the reaction
sequence when all the upstream reactions are in rapid equilibrium.  For
example, if $k_{-1}\gg k_2$, then the reaction between state 0 and
state 1 is maintained near equilibrium, and $\eta_2=J/J_-$ for the
reaction between state 1 and state 2.

Quantities such as $\frac{\partial\ln J}{\partial \ln c}$ have been
discussed in the context of allosteric regulation, by a metabolite
$c$, of some enzyme which in turn regulates the $J$ \cite{kacser}.
Brand and his coworkers have extensively used an empirical,
``top-down'' elasticity analysis in assessing the fractional changes
in the fluxes of metabolic reactions in response to a change in the
concentration of an effector \cite{brand}. These studies, which
emphasize the interactions between metabolites and enzymes, are
different from the above direct sensitivity of $J$ with respect to
an intermediate concentration due solely to the nature of
stoichiometric networks.

    In terms of the stoichiometric sensitivity, we now ask a
typical engineering question. How can one reduce (or increase) the
sensitivity of $c_1$ with respect to $J$, with given substrate
source $c_0$, flux $J$, intermediary metabolite concentration $c_1$?
We assume here that $c_1$ is a regulator and/or control agent for
some other parts of the cell, the specific nature of which is not of
our concern. To change $\eta_1$ according to Eq. \ref{eta1} while
maintaining $c_1$ at its fixed value, the effective $k_1$ and
$k_{-1}$ must be changed simultaneously so that
\begin{equation}
  \delta c_1 = \delta k_1 \left(\frac{\partial c_1}{\partial k_1}\right)
    + \delta k_{-1} \left(\frac{\partial c_1}{\partial
    k_{-1}}\right) = 0,
\end{equation}
which leads to
\begin{equation}
\frac{\delta k_1}{k_1} =
    \left(\frac{\delta k_{-1}}{k_{-1}}\right)
    \frac{c_1k_{-1}}{c_0k_1}.
\end{equation}
However, by altering enzyme activity, the effective $k_1$ and
$k_{-1}$ can be changed only according to $\delta k_1/k_1 = \delta
k_{-1}/k_{-1}$. Thus the sensitivity around a set-point flux and
intermediate concentration cannot be effectively adjusted by
increasing or reducing the enzyme activity for the reaction $c_0
\rightleftharpoons c_1$. To adjust $k_1$ and $k_{-1}$ independently
of one another it is necessary to change the effective $\Delta G^o$
= $RT\ln(k_{-1}/k_1)$ for reaction $0 \rightleftharpoons 1$. (Doing
this is known as impact control in \cite{kho}). From Eqs. (\ref{c1})
and (\ref{eta1}) it can be shown that $J=k_1c_0-k_{-1}c_1$
$=$ $J_-(k_1c_0/k_{-1}c_1-1)$, hence
\begin{equation}
    \eta_1= \frac{c_0}{c_1}e^{-\Delta G^o/RT}-1.
\label{eta1g0}
\end{equation}
Eq. \ref{eta1g0} is our first key result.  It shows that the
sensitivity of an intermediary metabolite concentration with respect
to the steady-state flux is governed by the $\Delta G^o$, the Gibbs
free energy of the intermediate formation. Fig. 1 shows how $\eta_1$
changes as a function of the $\Delta G^o$.  Qualitatively, this can
be understood without the mathematics: There are two extreme cases
of a steady state of the pathway in (\ref{rxn}).  Case I (far from
equilibrium) is that $J=k_1c_0-k_{-1}c_1$ $\gg$ $k_{-1}c_1$. In this
case $\Delta G^o/RT \ll \ln(c_0/c_1)$ and $\eta_1 \gg 1$. Case II
(near equilibrium) is that the $k_{-1}c_1$ $\gg$ $J$. In the case
where the reaction $0\rightleftharpoons1$ is maintained near
equilibrium, there is little sensitivity of $c_1$ to flux $J$. High
sensitivity occurs when $k_1 \gg k_{-1}$; all other things being
equal, the smaller the value of $k_{-1}$, the smaller the value of
$\Delta G^o$, and the greater the sensitivity. Note that $\eta_1$
goes to zero when $c_0$ and $c_1$ are in equilibrium; $\eta_1$
computed by Eq. \ref{eta1g0} remains positive for all cases for
which the reaction proceeds in the positive direction.

\begin{figure}[h]
\begin{center}
\includegraphics[height=7cm,angle=-90]{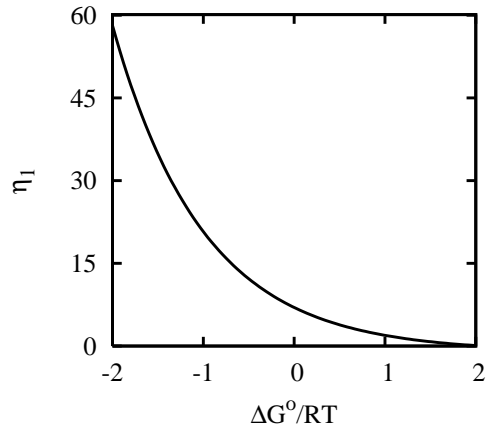}
\end{center}
\caption{Relation between Gibbs free energy, $\Delta G^o$ and
the stoichiometric sensitivity $\eta_1$.  For reaction
$0 \rightleftharpoons 1 \rightarrow$, with $c_0/c_1=8$.
Smaller the $\Delta G^o$, greater the sensitivity ($\eta_1$) of
$c_1$ in response to output flux  (Eq. \ref{eta1g0}).
Everything else being equal, greater
$\Delta G^o$ means greater backward flux from $c_1$ to $c_0$.
When it is significantly greater than the net flux $J$, the level
of $c_1$ will be insensitive to changes
in $J$.
}
\end{figure}

\subsection{Impact of Futile Cycles on Stoichiometric Sensitivity}

    For isolated chemical reactions, modifying
 $\Delta G^o$ can be accomplished only through modifying the solvent
conditions.  Such a mechanism is clearly not of primary importance
for cellular biochemical reactions.  Structural modifications change
the nature of the chemical reactions, leading to different values of
$\Delta G^o$. This mechanism represents a possible biological
strategy from an evolutionary standpoint, but it is not useful as a
mechanism for dynamic regulation of the sensitivity in a cell.

\begin{figure}[h]
\begin{center}
\includegraphics[height=7cm]{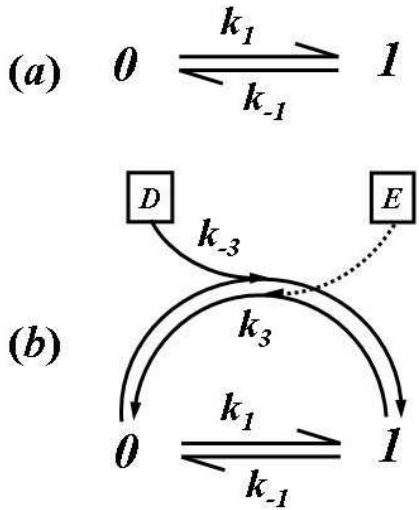}
\end{center}
\caption{Shown in ($a$), a biochemical reaction between
specie 0 and 1 in isolated system reaches its
equilibrium with concentrations $c_1^{eq}/c_0^{eq}=$
$k_1/k_{-1}=$ $e^{-\Delta G^o/RT}$.  Enzyme can change
the rate constants, but not the free energy difference
$\Delta G^o$.  However, if this reaction is coupled to
other reactions in an open biochemical network as shown
in ($b$), a futile cycle is able to shift the population
ratio $c_1/c_0$ to be greater (or less) than the
equilibrium value $k_1/k_{-1}$.  In ($b$), the
additional reactions involve species $D$ and $E$.  There
is now a futile cycle involving species 0 and 1.   The
equilibrium between $D$ and $E$ is $c_D^{eq}/c_E^{eq}=$
$c_1^{eq}k_3/(c_0^{eq}k_{-3})=$ $k_1k_3/(k_{-1}k_{-3})$.
If the concentrations of $D$ and $E$ are not at their
equilibrium, then $\ln(c_Ek_1k_3/(c_Dk_{-1}k_{-3}))=
\Delta G_{DE}\neq 0$, which is the active
energy source (e.g., nucleotide hydrolysis) that pumps
the futile cycle.  In a steady state this energy is
dissipated as heat.  The same mechanism is behind the
nuclear Overhauser effect in magnetic resonance, kinetic
proofreading in biosynethesis \cite{jjh}, and catalytic
wheel \cite{tsong}.
}
\end{figure}

    Since the free energy of formation of an intermediary
metabolite (at a given temperature, pressure, and solvent condition)
cannot be altered, is there a solution to reducing (increasing) the
sensitivity in a reaction network? One of the possible mechanisms
for increasing or reducing $\eta_1$ is a coupling between the
reaction and a futile cycle. Fig. 2 shows a futile cycle attached to
the reaction $0\rightleftharpoons1$, where if
$\frac{c_Dk_{-1}k_{-3}}{c_Ek_1k_3}$ $\neq$ 1, then the apparent free
energy difference between species $0$ and $1$,
$\Delta\widetilde{G}^o$ is
\begin{equation}
   e^{\Delta\widetilde{G}^o/RT} =
   \frac{k_{-1}+\hat{k}_3}{k_1+\hat{k}_{-3}}
    = e^{\Delta G^o/RT}
    \left(\frac{1+\sigma e^{\Delta G_{DE}/RT}}
                {1+\sigma}\right)
\end{equation}
in which $\sigma=\hat{k}_{-3}/k_1$, $\hat{k}_3=$ $k_3c_E$ and
$\hat{k}_{-3}=$ $k_{-3}c_D$ are pseudo-first order rate constants,
and $\Delta G_{DE}$ $=$
$RT\ln\frac{k_1\hat{k}_3}{k_{-1}\hat{k}_{-3}}$ is the chemical
driving force in the futile cycle in Fig. 2. By {\em apparent} free
energy, we mean that we would treat the nonequilibrium steady-state 
concentration ratio between $0$ and $1$ as they were in an 
equilibrium: $c_1/c_0 = e^{-\Delta\widetilde{G}^o/RT}$. In Eq.
\ref{eta1g0} the sensitivity $\eta_1$ is expressed as a function of
$\Delta G^o$. With the reaction coupled to the futile cycle
illustrated in Fig. 2, $\Delta G^o$ $\rightarrow$
$\Delta\widetilde{G}^o$, and $\eta_1$ is expressed as a function of
$\Delta\widetilde{G}^o$
\begin{equation}\label{eq.eta1g0p}
  \eta_1= \frac{c_0e^{-\Delta\widetilde{G}^o/RT}}{c_1}-1.
\end{equation}
Eq. \ref{eq.eta1g0p} can be expressed
\begin{equation}
   \eta_1 = \left(\eta_1^o+1\right)
        \left(\frac{1+\sigma }
        {1+\sigma e^{\Delta G_{DE}/RT}}\right)-1
\label{etavsdg}
\end{equation}
where $\eta_1^o$ is equal to $\eta_1$ in the absence of the futile
cycle ($\Delta G_{DE}=0$).  This is our second key result.  It shows
that the change in sensitivity $\eta_1$ can be controlled through
the amount of available energy, i.e. the driving force $\Delta
G_{DE}$ (also see Appendix B). The amount of energy consumed by each
turn of the futile cycle is $|\Delta G_{DE}|$.  Fig. 3 shows how
the sensitivity $\eta_1$ varies as functions of the energy
expenditure $\Delta G_{DE}$ for different $\sigma$, the relative
rates of the two steps in the loop.

\begin{figure}[h]
\begin{center}
\includegraphics[height=8cm, angle=-90]{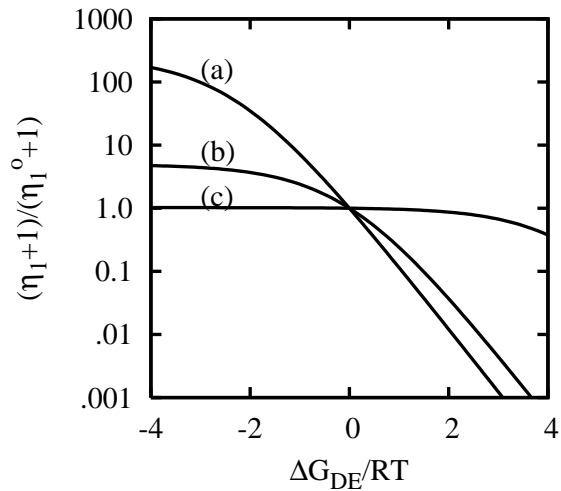}
\end{center}
\caption{Stoichiometric sensitivity ($\eta_1$) as a function of
the driving force in the futile cycle ($\Delta G_{DE}$) given in
Eq. (\ref{etavsdg}), with different values for the parameter
$\sigma$: (a) $\sigma=10$, (b) $\sigma=1$, and
(c) $\sigma=0.01$. $\eta_1^o$ is the corresponding $\eta_1$ when
$\Delta G_{DE}=0$, i.e., without the regulation from the futile cycle.
Note from Fig. 2, when the $\Delta G_{DE}<0$ the flux in the
futile cycle goes clockwise and when $\Delta G_{DE}>0$, it is counter
clockwise; They correspond to $\eta_1<\eta_1^o$ and
$\eta_1>\eta_1^o$, respectively.
}
\end{figure}

\begin{figure}[h]
\begin{center}
\includegraphics[height=6.7cm]{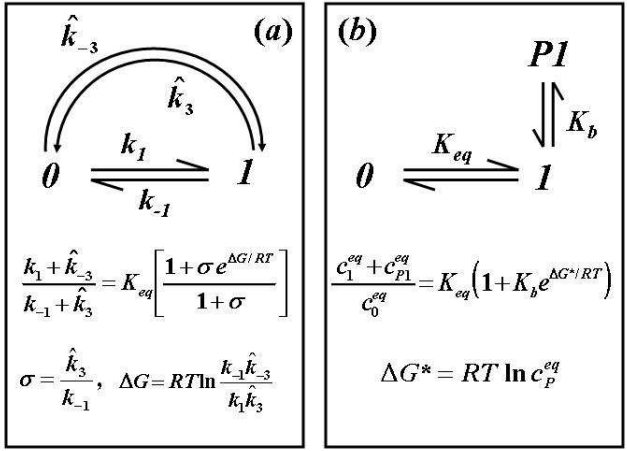}
\end{center}
\caption{Futile cycle ($a$) versus macromolecular binding
($b$), two mechanisms for modifying the equilibrium
concentration ratio between species $0$ and $1$. ($a$) The
futile cycle scheme requires active energy expenditure
$\Delta G$ but only small amount of proteins as the enzyme
for the additional reaction. $\hat{k}_{\pm 3}$
are pseudo-first order rate constants:
$\hat{k}_3=k_3c_E$ and $\hat{k}_{-3}=k_{-3}c_D$ in Fig. 2.
When $\Delta G=0$, the population ratio becomes $K_{eq}$.
($b$) The macromolecular binding scheme requires no
energy cost, but it requires a large amount of protein
being synthesized in advance.  When the concentration of
if the protein ($P$) $c_P^{eq}=0$, one has
$\Delta G^*=-\infty$ and the population ratio
becomes $K_{eq}$.
}
\end{figure}

A third key result follows from Eq. \ref{etavsdg}, which allows us
to diagnose the putative role of a futile cycle in a biological
system. When $\Delta G_{DE}<0$, then the futile cycle of
Fig. 2 is driven in the clockwise direction,
moving the reaction (at fixed $c_0$ and $c_1$) away from
equilibrium, and increasing the sensitivity of $c_1$ to changes in
flux. If the futile cycle is thermodynamically driven in the
counterclockwise direction, i.e., when $\Delta G_{DE}>0$, then
sensitivity of $c_1$ is reduced, i.e., the intermediate
concentration is made robust to changes in flux. In both cases
there is an energy expenditure and heat dissipation.  The
``high grade'' chemical energy is transformed into ``low grade''
heat, but is not wasted from the standpoint of information regulation.

\subsection{Sensitivity of Intermediate Concentration to Input Concentration}

To investigate the potential impact of a futile cycle on the
sensitivity of intermediate concentration to the input
concentration, we define $\zeta_1$ as the sensitivity of
intermediate concentration $c_1$ to changes in input concentration
$c_0$, at a given steady state flux $J$,
\begin{equation}
  \zeta_1 = \frac{\partial \ln c_1}{\partial \ln c_0}.
\end{equation}
From the equation $c_1$ $=$ $e^{-\Delta G^o/RT}(1-J/(k_1c_0))c_0$,
it is straightforward to show that
\begin{equation}
  \zeta_1^o = \left(\frac{c_0}{c_1}\right)e^{-\Delta G^o/RT},
\end{equation}
when there is no futile cycle acting on the reaction. For the case
when the futile cycle of Fig. 2 is present,
\begin{equation}\label{eq.zeta1}
 \zeta_1 = \left(\frac{c_0}{c_1}\right)e^{-\Delta \widetilde{G}^o/RT}
   = \zeta_1^o\left(\frac{1+\sigma}{1+\sigma e^{\Delta
   G_{DE}}}\right).
\end{equation}
Equation \ref{eq.zeta1} shows that the futile cycle has
qualitatively the same impact on intermediate concentration
sensitivity to input concentration as on intermediate concentration
sensitivity to flux.

\section{Optimal Intermediary Metabolite Sensitivity and Robustness}

    The previous discussion was based on the example of Eq. (\ref{rxn}).
We now present a more general theory for stoichiometric sensitivity
in biochemical networks.
For a given reaction $ r \rightleftharpoons p$ in a biochemical
network, its kinetics, according to the law of mass actions, 
and thermodynamics are determined by its forward and backward 
fluxes $J_+$ and $J_-$ \cite{my05bpc}:
\begin{equation}
    J = J_+-J_-, \hspace{0.3cm}
    \Delta G = RT\ln\frac{J_-}{J_+}.
\end{equation}
If we perturb concentrations and flux around a steady-state we have
\begin{eqnarray}
    \delta J &=& \delta J_+ -\delta J_-, \nonumber
\\
    \delta\Delta G &=& RT\left(\frac{\delta J_-}{J_-}
                -\frac{\delta J_+}{J_+}\right).
\label{ddG}
\end{eqnarray}
Solving $\delta J_+$ and $\delta J_-$ in terms of $\delta J$ and
$\delta\Delta G$, we have equations that relate changes in the
concentrations of the reactant and the product, $c_r$ and $c_p$ to
change in the metabolic flux and free energy:
\begin{eqnarray}\label{eq.dCdJ}
    \frac{\delta c_r}{c_r} &=& \frac{\delta J_+}{J_+}
    = \frac{\delta J}{J} + \frac{J_-}{J}
        \left(\frac{\delta\Delta G}{RT}\right), \nonumber
\\
    \frac{\delta c_p}{c_p} &=& \frac{\delta J_-}{J_-}
    = \frac{\delta J}{J} + \frac{J_+}{J}
        \left(\frac{\delta\Delta G}{RT}\right).
\end{eqnarray}
If the flux $J$ changes as the metabolic network moves from one
steady state to another, the concentrations $c_r$ and $c_p$ change
according to Eq. \ref{eq.dCdJ}.
The {\em total relative change} in reactant and product
concentrations associated with the reaction can be calculated as,
\[
  \left(\frac{\delta c_r}{c_r}\right)^2 +
  \left(\frac{\delta c_p}{c_p}\right)^2 =
    2\left(\frac{\delta J}{J}\right)^2
    +2\left(\frac{J_++J_-}{J}\right)
\]
\begin{equation}
    \left(\frac{\delta J}{J}\right)
    \left(\frac{\delta\Delta G}{RT}\right)
    +\frac{J_+^2+J_-^2}{J^2}
    \left(\frac{\delta\Delta G}{RT}\right)^2.
\end{equation}
We define total sensitivity as
\begin{eqnarray}
   \eta &=& \sqrt{
  \left(\frac{\partial\ln c_r}{\partial\ln J}\right)^2+
  \left(\frac{\partial\ln c_p}{\partial\ln J}\right)^2}
\nonumber\\
    &=& \sqrt{(1-J_-\theta)^2+(1-J_+\theta)^2},
\label{eta}
\end{eqnarray}
where $\theta$ $=$ $-\frac{1}{RT}\left(\frac{\delta\Delta G}{\delta
J}\right)$ is the steady-state change in the $\Delta G$ in response
to a change in $J$.  It can be thought of as the nonlinear
biochemical resistance of the reaction.  With given $J_+$ and $J_-$,
$\eta$ reaches its minimum $\eta^*=$
$\left(\frac{J^2}{J_+^2+J_-^2}\right)^{1/2}$ when $\theta$ is
$\theta^*=$ $\frac{J_++J_-}{J_+^2+J_-^2}$.  This is the least
sensitive, maximal robustness condition for the reaction,
irrespective of how the reaction is situated in a network. This
result suggests that in maintaining certain reactions in metabolic
pathways near equilibrium, i.e., $J\ll J_+,J_-$, biological systems
may tend to minimize the sensitivities of concentrations of certain
key species to perturbations in flux. This insight may be used as a
lead for identifying regulatory sites in metabolic systems.

This results allows us to associate chemical thermodynamics with
robustness in a biochemical network. However, as we have seen in the
previous section, it is not possible to increase or decrease
robustness by increasing or decreasing the enzyme activity for a
given reaction $r\rightleftharpoons p$. By increasing enzyme
activity, $J_+$, $J_-$, and $J$ will all increase in the same
proportion. Hence the minimum value of $\eta$ is not affected. A
futile cycle, however, is capable of regulating the minimal $\eta$.

Several special cases are important.

\noindent ({\em i}) If the reaction is near equilibrium, $J_+,J_-\gg
J$, then we have the approximate relationship between $\Delta G$ and $J$
\[
    \Delta G= RT\ln\left(1-\frac{J}{J_+}\right) \approx
    -RT\left(\frac{J}{J_+}+\frac{1}{2}\left(\frac{J}{J_+}\right)^2
        \right)
\]
Hence, $\theta$ $\approx$ $\frac{1}{J_+}(1+\frac{J}{J_+})$.
Similarly we have $\theta$ $\approx$
$\frac{1}{J_-}(1-\frac{J}{J_-})$.  Substitute these into Eq. (\ref{eta})
we have $\eta$ $\approx$ $\frac{\sqrt{2}\Delta G}{RT}$. In this
regime, the total sensitivity of the concentrations to the flux,
$\eta$, is simply proprotional to the driving force $\Delta G$. It
is clear that the sensitivity is related to how far the systems is
away from equilibrium.

\noindent ({\em ii}) If the concentration $c_p$ is clamped, then
the $\delta J_-=0$ in Eq. (\ref{ddG}) and $\delta\Delta G$ $=$
$-\frac{RT}{J_+}\delta J$.  Hence $\theta$ $=$
$\frac{1}{J_+}$, and $\eta$ $=$ $\frac{J}{J_+}$.  The ratio
$\frac{J}{J_+}$ is known as irreversibility of the reaction; for an
irreversible reaction it is unity, and it approaches zero as the
reaction approaches equilibrium.  This result was first obtained in
\cite{newsholme2}.

\noindent ({\em iii}) By a similar argument, if the $c_r$ is clamped
then $\theta$ $=$ $\frac{1}{J_-}$, and $\eta$ $=$ $\frac{J}{J_-}$.
This is our Eq. (\ref{eta1}).

\noindent ({\em iv}) If the reaction is a control point for the
flux, then the crossover theorem for unbranched reaction pathway
\cite{rolleston,cross} dictates that for
$\delta J>0$, one has $\delta c_r \ge 0$ and $\delta c_p \le 0$.
This yields $\frac{1}{J_+}$ $\le$ $\theta$ $\le$ $\frac{1}{J_-}$.
Substituting this into Eq. (\ref{eta}), one can easily show that
$\eta$ has upper and lower bounds
\begin{equation}
    \frac{J}{\sqrt{J_+^2+J_-^2}} \le \eta \le
        \frac{J}{J_-}.
\end{equation}

\section{Comparison of Two Mechanisms for Shifting Apparent Equilibria:
Macromolecular Binding and Futile Cycling}

    Our analysis of sensitivity regulation would not
be complete without discussing macromolecular binding as a mechanism
for shifting apparent equilibria \cite{wyman,schellman}. For
example, consider the case where a protein $P$ binds $p$ but not $r$
in the reaction $r \rightleftharpoons p$ with equilibrium constant
$K_{eq}$. In this case, although the presence of $P$ does not change
the equilibrium constant between $r$ and $p$, it does change the
equilibrium concentration ratio between total $p$ ($p$ plus $pP$)
and $r$ \cite{schellman}:
\begin{equation}
    \frac{c_p^{eq}+c_{pP}^{eq}}{c_r^{eq}} = \frac{c_p^{eq}}{c_r^{eq}}
        \left(1+\frac{c_{pP}^{eq}}{c_p^{eq}}\right)
             = K_{eq} \left(1+K_bc_P^{eq}\right),
\end{equation}
in which $K_b$ is the association equilibrium constant between $P$
and $p$.  Thus it is possible to reduce the sensitivity $\eta_p$ in
Eq. (\ref{eta1}) by introducing a binding protein for intermediary
metabolite $p$ (or increase $\eta_r$ with a binding protein for $r$)
that essentially serves as a buffer.
Hence, there exist at least two distinct mechanisms of modulating
robustness and sensitivity in biochemical systems. One is based on
allosteric binding, and another is based on reversible chemical
modification via futile cycle with energy expenditure. This issue
was raised as early as 1971 \cite{fischer} after the discovery of
enzyme regulation by reversible chemical modification in terms of
protein phosphorylation and dephosphorylation via a futile cycle
\cite{fischerkreb}. Specifically, Fischer asks \cite{fischer} ({\em
i}) ``Why have organisms found it advantageous to develop separate
mechanisms to control the activity of enzymes, namely, by
noncovalent (allosteric) changes in structure mediated by
appropriate effectors (binding), and by covalent modifications (via
futile cycle) of the proteins?'' and ({\em ii}) ``Why are these two
mechanisms, i.e., noncovalent allosteric regulation and covalent
modification via phosphorylation, usually superimposed on one
another even though the changes in conformation resulting in either
activation or inhibition are essentially the same?''

The essential difference between the binding mechanism and a futile
cycle is that reversible binding does not expend energy. Yet there
are important tradeoffs associated with macromolecular binding.
Specifically, buffering control by macromolecular binding requires
the effectors, say a protein, to be present in an amount
approximately equal to that of the metabolite. The acting principle
here is that significant binding is necessary to achieve a
significant shift in the effective equilibrium. The futile cycle
approach requires only a relatively small amount of enzyme to
catalyze the cyclic reactions. From a control systems perspective,
if feedback information is tied to the concentration of some
protein, then the former approach is ideal. On the other hand, if it
is necessary to achieve signal amplification where only a small
number of copies of a protein are activated, a
phosphorylation-dephosphorylation futile cycle is appropriate for
the task. In addition, the costs of binding regulation and the
futile cycles regulation are different. The former requires
significant amount of biosynthesis of effectors in advance, while
the latter requires only a small amount of enzymes for the
hydrolysis reaction. The latter consumes energy during the
regulation while the former pays in advance in the biosynthesis. In
engineering terms, this is an issue of material cost versus energy
utilization, an issue of overhead versus operational costs.

\section{Regulatory Sensitivities and Statistical Thermodynamics}

    The present work suggests an important relation between
the regulatory sensitivity and robustness of metabolic
systems in cell and the thermodynamics of biochemical reactions.
This rather unexpected quantitative relationship deserves further
investigation, especially from a systems biology perspective.
In this section, we provide some initial discussions for the
subject.

\subsection{Reversibility of biochemical reactions}

All chemical reactions are reversible, although in some cases the
backward rate may be negligibly small. Approximating such cases as
irreversible is often acceptable in kinetic analysis, but is
problematic for a thermodynamic analysis as illustrated in
\cite{my03bpc}. In the biochemical literature a reaction usually is
operationally considered irreversible if $J_+/J_-$ $>$ 5
\cite{rolleston}. When $J_+/J_-$ $\approx 1$, i.e., $J_+$, $J_-$
$\gg$ $J=J_+-J_-$, a reaction is considered to be near equilibrium
and there is a linear relation between its flux and the chemical
potential difference $\Delta G$ $=$ $RT\ln (J_+/J_-)$ $\approx$
$RTJ/J_-^{eq}$.  Thus the equilibrium forward and reverse flux
$J_-^{eq}$ $=$ $J_+^{eq}$ (divided by $RT$) is the conductance of a
biochemical reaction operating near equilibrium \cite{my02bj}.

\subsection{Sensitivity and concentration fluctuations}

    Because of the thermal agitations and the stochastic nature
of molecular reactions, the concentrations of a biochemical species
fluctuate, in a test tube reactions and in cells \cite{my04pnas}.
Rigorously, the concentrations (such as $c_1$ and $c_2$) discussed
so far represent the mean values of the concentrations of these
species. If one is able to measure the fluctuating concentration of
a species as a function of time, say $c_1(t)$ for species 1, then
one may calculate the magnitude of the concentration fluctuation by
$\int_0^{\infty} \langle \Delta c_1(0)\Delta c_1(t)\rangle dt$ where
$\Delta c_1(t) = c_1(t)-\langle c_1\rangle$ is the deviation of
$c_1(t)$ from its mean value. The notation $\langle \cdots\rangle$
denotes average of the quantity in the brackets, and
$\langle \Delta c_1(0)\Delta c_1(t)\rangle$ is known as
time-correlation function of fluctuating $c_1$ \cite{my04pnas}.

    One might imagine that a biochemical species with
high sensitivity of concentration will display similarly large
concentration fluctuations.  This is indeed the case. The
sensitivities introduced in Eqs. (\ref{eta1}) and (\ref{eta2}) are
intimately related to the steady-state concentration fluctuations in
the intermediary metabolites \cite{my04pnas}. To see this, we note
that from Eqs. (\ref{eta1}) and (\ref{eta2})
$\eta_1=J\left(c_0k_1-J\right)^{-1}$ and
$\eta_2=J\left(\frac{c_0k_1k_2}{k_2+k_{-1}}-J\right)^{-1}$,
respectively.

It is shown in Appendix C (Eqs. \ref{mm} and \ref{nn} and the discussion
there) that the $c_0k_1$ and $c_0k_1k_2/(k_2+k_{-1})$ can be expressed
as:
\begin{equation}
      \left(c_0k_1\right)^{-1} = \int_0^{\infty}
            \frac{\langle \Delta c_1(0)\Delta c_1(t)\rangle}
             {\langle c_1\rangle^2} dt,
\label{fluc1}
\end{equation}
and
\begin{equation}
      \left(\frac{c_0k_1k_2}{k_2+k_{-1}}\right)^{-1}
        = \int_0^{\infty}
            \frac{\langle \Delta c_2(0)\Delta c_2(t)\rangle}
             {\langle c_2\rangle^2} dt.
\label{fluc2}
\end{equation}
The right-hand-sides of Eqs. (\ref{fluc1}) and (\ref{fluc2})
are the concentration fluctuations.  Thus, we see that smaller
the fluctuations, smaller the sensitivities.  Elf et. al. 
have reached a similar conclusion based on a linear approximation 
of a nonlinear, stochastic biochemical kinetics \cite{elf}.

\subsection{Futile cycle and heat dissipation}

 While the major function of
futile cycles in signal transduction seems to be improving the
performance of information processing against noise, the functional
roles of futile cycles in metabolic systems potentially include
improving sensitivity or robustness of metabolite concentrations. In
addition to affecting sensitivity, it is possible that futile cycles
play an important role in generating heart and regulating
temperature \cite{heinrich}.  Hence it is important to analyze
metabolic regulations with a systems perspective. In particular the
intriguing suggestion \cite{newsholme2} that the futile cycles are
important components in obesity, in weight loss, and even in the
so-called Atkins' diet \cite{obes,diet,natmed}, deserves further
investigation.  It is timely to rethink the issues of nutrition,
thermogenesis \cite{thermogenesis}, and futile cycles with a
molecular as well as modern systems biology and metabolic
engineering approach \cite{bailey,hanson,wp}.  The present work
provides a thermodynamic basis for studying futile cycles, which is
likely to be essential in such studies.


\section{Summary}

    Spending energy to gain control is not a foreign
concept in engineering. Since this strategy is hallmark of control
engineering, it should not come as surprise that biological cells
use energy in controlling metabolism, transcription, and
translation.  Thus futile cycles which utilize biochemical energy
are not necessarily ``futile''; they likely serve as mechanisms of
biochemical regulation.

It has become increasingly clear from our recent work that futile
cycles play a unique and essential role in cellular regulation and
signal transduction in the form of protein phosphorylation
dephosphorylation (and GTPase). While the phosphate group serves as
a structural signal in enzyme activation, the phosphorylation
reaction also provides a source of energy. The energy expenditure in
fact increases the accuracy \cite{my02traffic}, sensitivity
\cite{my03bpc}, specificity \cite{jjh}, and robustness
\cite{aoping1,my05prl} of the cellular information processing,
overcoming cellular internal noises from thermal fluctuations, small
copy numbers, and limited affinities.

However to date, the role of futile cycles in metabolic regulation
has been less quantitatively understood. In this work we have shown
that metabolic futile cycles shift the effective equilibrium
constants for biochemical reactions, modulating the sensitivity and
robustness of intermediate concentrations to changes in flux. By
shifting the effective equilibrium so that a reaction is moved away
from equilibrium, the stoichiometric sensitivity is increased.
Shifting it in the other direction reduces the sensitivity and
enhances robustness. The direction of the shift, and hence a
putative physiological role for a given futile cycle, can be
diagnosed from the direction of operation of the futile cycle. When
a futile cycle drives a reaction in the forward direction (clockwise
in Figs. 2b and 4a), then the sensitivity of the concentration of
intermediate species 1 to the steady-state flux is enhanced. When a
futile cycle drives the reaction in the direction opposite the net
flux (counterclockwise in Figs. 2b and 4a), then the sensitivity is
reduced and the intermediate concentration is made robust to changes
in the flux.

\section{Acknowledgments}


We thank Dr. Jim Bassingthwaighte for helpful comments and 
Dr. David Moskowitz (GenoMed Inc.) for pointing us to the
reference \cite{newsholme}.
This work was supported by NIH grants GM068610 and HL072011.

\section{Appendices}

\subsection{A: Sensitivity with different output flux controls}

Consider the reaction
\begin{equation}\label{rxnA}
 0 \overset{k_1}{\underset{k_{-1}}\rightleftharpoons}
  1 \overset{k_2}{\underset{k_{-2}}\rightleftharpoons}
  2 \overset{J}{\longrightarrow}
\end{equation}
With $c_0$ fixed, there are several ways to control the output flux
$J$, either ($i$) directly controlling flux $J$,
or ($ii$) by controlling the rate constant $k_3$ in the downsteam
reaction, $2 \overset{k_3}{\rightarrow} 3$, or
($iii$) by controlling the concentration of downstream species,
$c_3$ in $2\overset{k_3}{\underset{k_{-3}}\rightleftharpoons} 3$,
or the enzyme for the reaction.
We show here that all these three cases yield identical
expressions for the stoichiometric sensitivity.

    The kinetic equations for reaction system in (\ref{rxnA}) are
\begin{eqnarray}
    \frac{dc_1}{dt} &=& k_1c_0-(k_{-1}+k_2)c_1+k_{-2}c_2,
\\
    \frac{dc_2}{dt} &=& k_2c_1 -k_{-2}c_2-\left\{
        \begin{array}{l} J  \\
                 k_3c_2 \\
                 k_3c_2+k_{-3}c_3
        \end{array}\right.
\label{3case}
\end{eqnarray}
where the last term in Eq. (\ref{3case}) represents the three cases
above.

    Consider the dynamic equation for $c_2$ with the general
expression,
\begin{eqnarray}
    \frac{dc_1}{dt} &=& k_1c_0-(k_{-1}+k_2)c_1+k_{-2}c_2,
\\
    \frac{dc_2}{dt} &=& k_2c_1 -k_{-2}c_2-
            J(c_2,c_3,k_3,k_{-3}).
\end{eqnarray}
In steady-state, we have
\begin{eqnarray}
    k_1c_0-(k_{-1}+k_2)c_1+k_{-2}c_2 &=& 0,
\\
    k_2c_1 -k_{-2}c_2-J(c_2,c_3,k_3,k_{-3}) &=& 0.
\end{eqnarray}
Therefore, we have
\begin{eqnarray}
    -(k_{-1}+k_2) \left(\delta c_1\right)
    +k_{-2}\left(\delta c_2\right) &=& 0,
\\
    k_2\left(\delta c_1\right)
    -k_{-2}\left(\delta c_2\right)
    -\delta J(c_2,c_3,k_3,k_{-3}) &=& 0.
\end{eqnarray}
Hence, no matter how $J$ is changed, by $k_3$, by $c_3$, or $k_{-3}$,
as long as it is not by $c_2$, we have
\begin{eqnarray}
    -(k_{-1}+k_2) \left(\frac{\partial c_1}{\partial J}\right)
    +k_{-2}\left(\frac{\partial c_2}{\partial J}\right) &=& 0,
\\
    k_2\left(\frac{\partial c_1}{\partial J}\right)
    -k_{-2}\left(\frac{\partial c_2}{\partial J}\right)
    &=& 1.
\end{eqnarray}
Solving this pair of albegraic equations yields
\begin{equation}
        \frac{\partial c_1}{\partial J}
                = -\frac{1}{k_{-1}}, \hspace{0.3cm}
        \frac{\partial c_2}{\partial J}
                = -\frac{k_2+k_{-1}}{k_{-1}k_{-2}}.
\end{equation}

\subsection{B: A simple example of how futile cycle drives
concentration ratio away from equilibrium}

This example is motivated by the classic work on
kinetic proofreading \cite{jjh}.  Let's consider a 3-state kinetic
system with $A$, $B$ and $C$ shown in Fig. 5.  The
equilibrium constant $K_{12} = \frac{[B]^{eq}}{[A]^{eq}}$, and
reaction between $B$ and $C$ is coupled to an energy source
$\Delta G_{DE}=$ $RT\ln\frac{[E]k_{-2}^ok_{-3}}{[D]k_2^ok_3K_{12}}$.
We shall denote pseudo-first order rate constants $k_2=k_2^o[D]$ and
$k_{-2}=k_{-2}^o[E]$, and $\gamma=$
$\frac{K_{12}k_2k_3}{k_{-2}k_{-3}}$ $=e^{-\Delta G_{DE}/RT}$.
Then we have
\begin{equation}
    \frac{[C]}{[A]} = \frac{\left(1+K_{12}\right)[C]}{([A]+[B])}
              = \frac{k_{-3}}{k_3}
        \left(\frac{k_{-2}\gamma+k_3}{k_3+k_{-2}}\right)
\end{equation}
where $RT\ln\gamma=-\Delta G_{DE}$ is the amount of energy pumped input to the reaction.
$\gamma=1$ for a closed system in equilibrium.  For large $\gamma$, i.e.,
the kinetic cycle goes clockwise, it is possible to
have $\gamma k_{-2} \gg k_3 \gg k_{-2}$ and,
\begin{equation}
    \frac{[C]}{[A]} \approx \frac{k_{-3}}{k_3}
        \left(\frac{k_{-2}\gamma}{k_3}\right)
        \gg \frac{k_{-3}}{k_3},
\end{equation}
the expected equilibrium ratio between $B$ and $A_1$.  On the
other hand, for small $\gamma$, i.e., the energy pumping is
counter-clockwise, it is possible to have
$\gamma k_{-2} \ll k_3 \ll k_{-2}$, and
\begin{equation}
    \frac{[C]}{[A]} \approx  \frac{k_{-3}}{k_3}
        \left(\frac{k_3}{k_{-2}}\right)
        \ll \frac{k_{-3}}{k_3}.
\end{equation}

\subsection{C: Concentration fluctuations in open systems}

Here we consider concentration fluctuations in the kinetic pathway
with constant source $c_0$ and sink $c_3=0$:
\begin{equation}
  0 \overset{k_1}{\underset{k_{-1}}\rightleftharpoons}
  1 \overset{k_2}{\underset{k_{-2}}\rightleftharpoons}
  2 \overset{k_3}{\longrightarrow} 3.
\end{equation}
For a general theory on open, linear biochemical networks see
\cite{wjh-hq}.  In stochastic terms, the probability of the numbers
of species 1 and 2 being $m$ and $n$ at time $t$, $P(m,n,t)$
satisfies the chemical master equation \cite{gardiner,keizer}
\[
  \frac{dp(m,n,t)}{dt} = -\left(k_1c_0+m(k_{-1}+k_2)
	+(k_{-2}+k_3)n\right) p(m,n)
\]
\[
        +k_1c_0p(m-1,n)
    +(m+1)k_{-1}p(m+1,n)+(m+1)k_2p(m+1,n-1)
\]
\begin{equation}
    +(n+1)k_{-2}p(m-1,n+1)+(n+1)k_3p(m,n+1)
\label{master}
\end{equation}
From Eq. (\ref{master}) it is easy to show that the mean values of
$m$ and $n$ follow the standard deterministic kinetic equations
\begin{eqnarray}
    \frac{d}{dt}\langle m\rangle &=& k_1c_0
        -(k_{-1}+k_2)\langle m\rangle
        + k_{-2}\langle n\rangle
\\
    \frac{d}{dt}\langle n\rangle &=& k_2\langle m\rangle
        -(k_{-2}+k_3)\langle n\rangle
\end{eqnarray}
and furthermore their variances and covariance,
\begin{eqnarray}
    \frac{d}{dt}\langle(\Delta m)^2\rangle &=& k_1c_0
    +(k_{-1}+k_2)\langle m\rangle+k_{-2}\langle n\rangle
\nonumber\\
    &&-2(k_{-1}+k_2)\langle(\Delta m)^2\rangle
    +2k_{-2}\langle\Delta m\Delta n\rangle,
\nonumber\\
    \frac{d}{dt}\langle(\Delta n)^2\rangle &=& k_2\langle m\rangle
        +\left(k_{-2}+k_3\right)\langle n\rangle
\nonumber\\
       && +2k_2\langle \Delta m\Delta n\rangle
        -2(k_{-2}+k_3)\langle (\Delta n)^2\rangle,
\nonumber
\\
    \frac{d}{dt}\langle\Delta m\Delta n\rangle &=&-k_2\langle m\rangle
        -k_{-2}\langle n\rangle
        +k_2\langle(\Delta m)^2\rangle
        +k_{-2}\langle (\Delta n)^2\rangle
\nonumber\\
        &&-(k_{-1}+k_2+k_{-2}+k_3)\langle \Delta m\Delta n\rangle,
\nonumber
\end{eqnarray}
where $\Delta m$ $\triangleq$ $m-\langle m\rangle^{ss}$ and $\Delta
n$ $\triangleq$ $n-\langle n\rangle^{ss}$. The steady state mean
values
\begin{equation}
    \langle m \rangle^{ss} = \frac{k_1(k_{-2}+k_3)c_0}
            {k_{-1}(k_{-2}+k_3)+k_2k_3},
    \langle n \rangle^{ss} = \frac{k_1k_2c_0}
            {k_{-1}(k_{-2}+k_3)+k_2k_3},
\end{equation}
which agree with Eqs. (\ref{c1}) and (\ref{c2}) if we substitute
$J=\langle n\rangle^{ss}k_3$, and more,
\begin{equation}
       \langle(\Delta m)^2\rangle^{ss} = \langle m\rangle^{ss},
        \hspace{0.2cm}
       \langle(\Delta n)^2\rangle^{ss} = \langle n\rangle^{ss},
        \hspace{0.2cm}
       \langle \Delta m \Delta n\rangle^{ss} = 0.
\label{meanvar}
\end{equation}
In fact, one can verify that the stationary probability distribution
for $m$ and $n$ is Poissonian \cite{wjh-hq}. $\frac{\left(\langle
m\rangle^{ss}\right)^m}{m!}e^{\langle m\rangle^{ss}}$
$\frac{\left(\langle n\rangle^{ss}\right)^n}{n!}e^{\langle
n\rangle^{ss}}$ is the stationary distribution for Eq.
(\ref{master}). However, while Eq. (\ref{meanvar}) indicates that
the steady-state fluctuations for $m$ and $n$ are independent.  It
does not mean that $m$ and $n$ are compleltely independent. In fact,
we have $\langle \Delta m(0) \Delta n(t)\rangle$ $\neq$ 0 for $t\neq
0$:
\[
\langle \Delta m(0) \Delta m(t)\rangle = \hspace{2in}
\]
\begin{equation}
    \frac{\langle  m\rangle^{ss}}{\lambda_1-\lambda_2}
    \left(-(\lambda_2+k_{-1}+k_2)e^{\lambda_1 t}+
     (\lambda_1+k_{-1}+k_2)e^{\lambda_2 t}\right)
\label{mm}
\end{equation}
\[
\langle \Delta n(0) \Delta n(t)\rangle = \hspace{2in}
\]
\begin{equation}
    \frac{\langle  n\rangle^{ss}}{\lambda_1-\lambda_2}
    \left(-(\lambda_2+k_{-2}+k_3)e^{\lambda_1 t}+
     (\lambda_1+k_{-2}+k_3)e^{\lambda_2 t}\right)
\label{nn}
\end{equation}
\begin{equation}
\langle \Delta m(0) \Delta n(t)\rangle =
    \frac{k_2\langle  m\rangle^{ss}}{\lambda_1-\lambda_2}
    \left(e^{\lambda_1 t}-e^{\lambda_2 t}\right)
\end{equation}
\begin{equation}
\langle \Delta n(0) \Delta m(t)\rangle =
    \frac{k_{-2}\langle  n\rangle^{ss}}{\lambda_1-\lambda_2}
    \left(e^{\lambda_1 t}-e^{\lambda_2 t}\right)
\end{equation}
where $\lambda_1$ and $\lambda_2$ are the two eigenvalues of the
matrix
\begin{equation}
    \MF = \left(\begin{array}{cc}
        -k_{-1}-k_2 & k_{-2} \\
        k_2 & -k_{-2}-k_3
    \end{array}\right).
\label{mf}
\end{equation}
Note that the concentation $c_1$ ($c_2$) and number of
molecules $m$ ($n$) differ by the volume of the system.
Hence, dentifying
$\frac{\langle c_1(0)c_1(t)\rangle}{\langle c_1\rangle^2}$ $=$
$\frac{\langle m(0)m(t)\rangle}{(\langle m \rangle^{ss})^2}$ and
$\frac{\langle c_2(0)c_2(t)\rangle}{\langle c_2\rangle^2}$ $=$
$\frac{\langle n(0)n(t)\rangle}{(\langle n \rangle^{ss})^2}$, and
integrating Eqs. \ref{mm} and \ref{nn} we arrive at
Eqs. \ref{fluc1} and \ref{fluc2}.

    We shall also point out that the stochastic dynamics on
a mesoscopic scale can be described by a Fokker-Planck
equation \cite{gardiner,keizer}
\begin{equation}
    \frac{\partial}{\partial t}P(m,n,t) =
\nabla\cdot\left[\frac{\MGamma}{2}\nabla P-\MF{\Delta m\choose
\Delta n}P\right]
\end{equation}
in which
\[
    \MGamma(m,n) = \hspace{2in}
\]
\begin{equation}
\left(\begin{array}{cc}
        k_1c_0+(k_{-1}+k_2)m+k_{-2}n & -k_2m-k_{-2}n
\\
        -k_2m-k_{-2}n & k_2m+(k_{-2}+k_3)n
    \end{array}\right)
\end{equation}
and $\MF$ is given in Eq. (\ref{mf}). We see that $\MGamma$ can be
expressed as $\frac{1}{2}\MS(\vJ^++\vJ^-)\MS^T$ where the
stoichiometric matrix \cite{my05bpc}
\begin{equation}
       \MS = \left(\begin{array}{ccc}
        -1& +1& 0 \\ 0& -1& +1  \end{array}\right)
\end{equation}
representing two species and three reactions in Eq. (\ref{rxn}).
According to the Fokker-Planck equation, the stationary distribution
$P^{ss}(m,n)$ is Gaussian centered arround $(\langle
m\rangle^{ss},\langle n\rangle^{ss})$ with covariant matrix
$\Msigma$ satisfying \cite{keizer,myprsa,aoping2}
\begin{equation}
        \MF\Msigma+\Msigma\MF^T
    = -\MGamma\left(\langle m\rangle^{ss},\langle n\rangle ^{ss}\right).
\label{sfdr}
\end{equation}
It can be easily verified via matrix multiplication that the
solution to the Eq. (\ref{sfdr}) is precisely Eq. (\ref{meanvar}).
We should point out that the fluctuation covariance is not
$\MF\MF^T$ as recently suggested \cite{orrell}. The information on
the concentration fluctuations is not contained in the relaxation
rate ($\MF$) alone; it has to depend on $\MGamma$ which represents
the {\em rate of fluctuations} \cite{my05bpc}.

\begin{figure}
\begin{center}
\includegraphics[width=1.8in,height=3in]{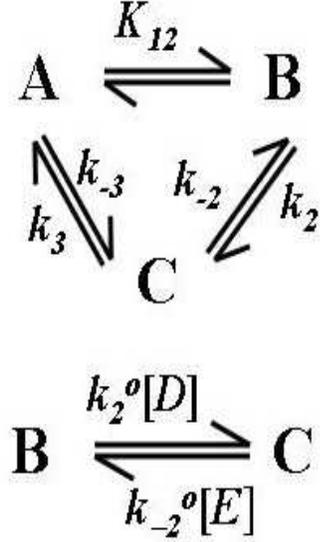}
\end{center}
\caption{A realistic mechanism, motivated by
the kinetic proofreading theory for protein
biosynthesis \cite{jjh}, showing how a driven
cyclic reaction, i.e., futile cycle, regulates
the population ratio between $A$ and $C$ away
from their equilibrium ratio $k_{-3}/k_3$.
The chemical driving force for the cycle is
$-\Delta G_{DE}=$
$RT\ln(K_{12}k_2^ok_3[D]/(k_{-2}^ok_{-3}[E]))$.
}
\end{figure}

\newpage

\end{document}